# Influence of deformation and annealing twinning on the microstructure and texture evolution of face-centered cubic high-entropy alloys

Christian Haase[1*], Luis A. Barrales-Mora[2]

[1]Steel Institute, RWTH Aachen University, Aachen, 52072 Germany

[2]Georgia Institute of Technology—CNRS 2958, George W. Woodruff School of Mechanical

Engineering, Metz Technopole, 2 Rue Marconi, 57070 Metz, France

## Abstract

The influence of the physical mechanisms activated during deformation and annealing on the microstructure and texture evolution as well as on the mechanical properties in the equiatomic CoCrFeMnNi high-entropy alloy (HEA) were investigated. A combination of cold rolling and annealing was used to investigate the HEA in the deformed, recovered, partially recrystallized, and fully recrystallized states. Detailed microstructure and texture analysis was performed by electron backscatter diffraction, transmission electron microscopy, and X-ray diffraction. The mechanical properties were evaluated using uniaxial tensile testing. A specific focus of this investigation was put on studying the influence of deformation and annealing twinning on the material behavior. It was substantiated that during cold rolling deformation, twinning facilitates the transition from the Cu-type to the Brass-type texture, whereas annealing-twinning leads to a strong modification of the texture formed during recrystallization. The formation of specific twin orientations and the randomization of the recrystallization texture were proven by experiments as well as by cellular automaton simulations. During tension of the cold-rolled and annealed material high work-hardenability was observed. We attribute this behavior primarily to the dominance of planar dislocation slip and reduced tendency for dynamic recovery, since deformation twinning was observed to activate only in the non-recrystallized grains and thus, contributed minimally to the overall plasticity. The correlation between deformation/annealing twinning and the material behavior was discussed in detail.

---
[*]Corresponding author. Tel.: +49 241 8095821; fax: +49 241 8092253. E-mail address:
*christian.haase@iehk.rwth-aachen.de* (Christian Haase).

# 1. Introduction

High entropy alloys (HEAs) are a new family of metallic materials, which are characterized by the alloying of more than 4 elements in usually equiatomic or near-equiatomic compositions. Despite the complex chemical composition of these alloys, they crystallize in simple crystal structures and form non-ordered solid solutions. The reason for this unique behavior and the idea behind HEAs is that significantly high mixing entropies increase the stability of the solid solution against the formation of intermetallic compounds [1]. These alloys were first introduced in separate publications by Cantor et al. [2] and Yeh et al. [3] and have rapidly gained the interest of the materials science community owing among others to their exceptional mechanical properties. In fact, HEAs are characterized by unusual combinations of properties not found in conventional alloys, such as high hardness, strength, and wear resistance, exceptional fracture toughness at cryogenic temperatures, and high corrosion and oxidation resistance [4-6]. The cause of these exceptional properties stems from different effects that influence the response of the materials to external stimuli. Besides the high-entropy effect that stabilizes the solid solution, also inherently slow diffusion kinetics, a severe lattice distortion, and the so-called 'cocktail effect' influence the properties of HEAs [5, 7-9], although the influence of these effects are discussed critically in open literature [10-12].

As in any other alloy, however, the properties are predominantly determined by the microstructure. For this reason, a major field of research has been dedicated to study microstructure and texture evolution of HEAs during their processing. HEAs have been processed by conventional thermomechanical treatment [13-15], severe plastic deformation [16, 17], and additive manufacturing [18, 19]. Even as single-phase materials, a variety of mechanical properties have been observed by varying the microstructure especially the grain size and the texture of HEAs [20-24]. In most of these investigations, the Co-Cr-Fe-Mn-Ni system, especially the prototypical equiatomic Cantor alloy (CrMnFeCoNi), has been utilized.

This alloy possesses a face-centered cubic (fcc) crystal structure, it has a moderately low stacking-fault-energy (SFE) and exhibits characteristics similar to low SFE metals, such as brass and high-manganese steel [13, 25]. Deformation in this and similar HEAs proceeds primarily by dislocation glide. However, at high tensile strains [26], during cryo and room temperature rolling [21], and under severe plastic deformation conditions [16], deformation twinning can be activated as an auxiliary mechanism leading to twinning-induced plasticity (TWIP). Furthermore, by variation of the chemical composition and by lowering the stability of the face-centered cubic (fcc) phase against the hexagonal phase, transformation-induced plasticity (TRIP) was observed in a non-equiatomic $Fe_{40}Mn_{40}Co_{10}Cr_{10}$ alloy [27].

In quaternary and quinary fcc HEAs, weak Brass-type textures have been observed even after 90% rolling at room and cryogenic temperatures [14, 28]. Regarding the microstructure, the material has showed at low deformation degrees slip lines and grain fragmentation and at high degrees also deformation-induced twins and shear bands. Upon recrystallization annealing, the texture has remained mostly weak but showing the presence of characteristic rolling texture components, such as the {110}<112> Brass, {110}<100> Goss, and {123}<634> S texture components [13]. In turn, the microstructure after recrystallization was found, depending on the annealing temperature, to consist of small equiaxed grains with the presence of recrystallization twins [29]. It is noted that it has been possible to refine substantially the microstructure by the application of conventional heat treatments resulting in an improvement of the mechanical properties [15, 22].

Evidently, the key for tailoring the mechanical properties is the understanding of the physical processes of microstructure and texture modification. The purpose of the present study is to investigate the role of the deformation and nucleation mechanisms on the texture and microstructure evolution in the equiatomic Cantor alloy during its processing by cold rolling and subsequent heat treatment. More specifically, the main focus of this work was put on revealing the influence of both deformation and annealing twinning on the material behavior. In addition, the relationship between microstructure and final mechanical properties is thoroughly discussed.

## 2. Applied methods

### 2.1. *Material and processing*

The equiatomic CoCrFeMnNi high-entropy alloy was investigated in this work. The exact chemical composition is given in Table 1. The corresponding SFE of the alloy at room temperature was estimated to be in the range between 18.3 and 27.3 mJ/m$^2$ [25, 30].

The alloy was produced as a 500 g-ingot using induction melting in Ar atmosphere followed by hot rolling at 1000 °C and homogenization at 1000 °C for 1 h. The hot-rolled sheet (3.1 mm) was further cold rolled up to 50% thickness reduction (1.55 mm). To investigate the influence of additional heat treatment on the material behavior, the cold-rolled sheets were annealed in the range between 500 °C and 900 °C for 1 h in an air furnace.

### 2.2. *Sample preparation and characterization techniques*

Specimens with the dimensions of 12 mm x 10 mm in the rolling (RD) and transverse direction (TD), respectively, were fabricated from the hot-rolled, cold-rolled and annealed sheets using electrical discharge machining. The samples were mechanically ground up to 4000 SiC grit paper followed by mechanical polishing using 3 µm and 1 µm diamond suspension. For X-ray diffraction (XRD) pole figure measurements, the middle layer of the RD-TD section was polished electrolytically at room temperature for 40 s at 24 V, whereas the RD-ND (ND - normal direction) section was electropolished for scanning electron backscatter diffraction (EBSD) using the same parameters as before. The electrolyte used for XRD and EBSD sample preparation consisted of 700 ml ethanol ($C_2H_5OH$), 100 ml butyl glycol ($C_6H_{14}O_2$), and 78 ml perchloric acid (60%) ($ClO_4$). Transmission electron microscopy (TEM) samples (~100 µm thick, 3 mm in diameter) were prepared using the same electrolyte as for XRD and EBSD samples in a double jet Tenupol-5 electrolytic polisher with a voltage of 22-24 V and a flow rate of 10 at 4-6 °C.

EBSD analyses were performed in a LEO 1530 field emission gun scanning electron microscope (FEG-SEM) operated at 20 kV accelerating voltage and a working distance of 10 mm. The HKL Channel 5 software was utilized to visualize the ESBD data as well as for data post-processing, removal of wild spikes and non-indexed points, taking at least five neighbor points into account. Subdivision of EBSD mappings into subsets containing only non-recrystallized (non-RX) or recrystallized (RX) grains was realized using an algorithm of the MATLAB®-based toolbox *MTEX* [31-33], as described in [34, 35]. The internal grain/subgrain misorientation was calculated based on the grain reference orientation (GROD-AO) value, which takes the average grain/subgrain orientation as a reference. The misorientation threshold value for subdivision was chosen as RX < 1.5° < non-RX. Grains containing less than 10 EBSD data points were disregarded. MTEX was also utilized to calculate the microtexture orientation distribution functions (ODFs). TEM analyses were performed in a FEI Tecnai F20 TEM operated at 200 kV.

X-ray pole figures were acquired utilizing a Bruker D8 Advance diffractometer, equipped with a HI-STAR area detector, operating at 30 kV and 25 mA, using filtered iron radiation and polycapillary focusing optics. In order to characterize the crystallographic texture, three incomplete (0–85°) pole figures {111}, {200}, and {220} were measured. The macrotexture ODFs were also calculated and visualized using *MTEX*. The volume fractions of the corresponding texture components were calculated using a spread of 15° from their ideal orientation. The overall intensity of the textures was characterized by the respective texture index $T$ which was calculated as [36]:

$$T = \oint [f(g)]^2 dg \tag{1}$$

where *f(g)* is the orientation density function and *g* denotes the orientation defined by the three Euler angles $g = (\varphi_1, \Phi, \varphi_2)$.

Mechanical properties were evaluated by uniaxial tensile tests at room temperature and a constant strain rate of $2.5 \times 10^{-3}$ s$^{-1}$ along the previous rolling direction on a screw-driven Zwick Z100 mechanical testing device. Flat bar tension specimens were used with a gauge length of 13 mm, gauge width of 2 mm, fillet radius of 1 mm, and a total length of 33 mm.

*2.3 Computer Simulations*

To study the formation of the recrystallization texture, computer simulations were performed by means of a parallel cellular automaton (CA) [37-39]. This model considers a front-tracking approach [37], where recrystallization proceeds gradually by grain boundary migration. The CA model is essentially a pure growth model that requires a separate nucleation model. In this contribution, the effect of two specific nucleation mechanisms, grain boundary and recrystallization twinning, were studied. At grain boundaries, the nuclei have been observed to possess orientations that are slightly misoriented, typically 5°-10°, to that of parent grains [38]. For this reason, in the simulations the orientation of the nuclei placed at grain boundaries was deviated from that of the parent grain by a value randomly chosen between 5°-10°. For nuclei inside the grains (bulk mechanism), two approaches were considered to approximate the effect of twinning. In the first approach, the deformation texture was rotated by 60°<111> (first order/Σ3-orientation relationship) and then disaggregated in single orientations. These orientations were used to initialize nuclei inside the grains. In the second approach, the orientations of these grains were sampled from those determined experimentally (Fig. 14). Afterwards, simulations with different relative number densities of nuclei from the two mechanisms were performed [40].

# 3. Results

3.1. *Material behavior during cold rolling*

The initial state of the alloy after warm rolling was single-phase face-centered cubic with a small amount of non-metallic inclusions formed during the casting process. The microstructure consisted of globulitic grains with a mean grain diameter of ~26.1 µm (including twin boundaries).

Figure 1 illustrates the microstructure evolution during cold rolling to thickness reduction levels of 10% to 50% by EBSD analyses. At the lowest degree of deformation (10%), grains were only slightly compressed in the ND and elongated along the RD (Fig. 1a). After 25% thickness reduction, grains were further elongated, and longitudinal microstructural deformation features were frequently observed. These features contained mainly slip bands and deformation twins in suitably oriented grains, as shown in Fig. 1b. At the highest degree of deformation (50%), the dislocation density increased substantially, as evidenced by the low indexing rate and high ingrain orientation gradients (Fig. 1c). In addition, micro shear bands were formed due to localized strain concentrations with an inclination of about 35-40° to the rolling plane. Macro shear bands were not observed. Fig. 1d reveals that deformation twins were developed with a high density. However, it must be noted that only twin bundles were successfully indexed as twins, since the resolution limit of conventional EBSD does not allow identification of individual nanotwins. TEM bright-field images in Fig. 2 allow more detailed analysis of the microstructure after 50% cold rolling. The main microstructural feature after 50% thickness reduction were dislocation cells with very high dislocation densities within the cell boundaries, as exemplarily shown in Fig. 2a. In addition, the deformation twins appeared parallel to each other in thicker bundles (Fig. 2b and c) and contained a high dislocation density (inset in Fig. 2c). The average twin thickness and spacing were in the range of 60 nm.

The evolution of the macrotexture is illustrated by constant $\varphi_2$ sections of the ODF and the corresponding calculated volume fractions of the main texture components in Figs. 4 and 5, respectively. A schematic illustration of the position of the main texture components formed during rolling of fcc metals is shown in Fig. 3. Their definitions are given in Table 2. The initial texture of the material after warm rolling was very weak with a texture index of 1.26 and constituted a high volume fraction of randomly oriented grains[‡]. With increasing degree of thickness reduction the texture transformed gradually from the initially random texture to an fcc rolling texture. The fraction of randomly oriented grains decreased successively and indicated the rotation of grains towards the ideal rolling texture components. However, the texture index after 50% thickness reduction was still weak with a value of 2.22. More specifically, the {001}<100> Cube texture component decreased in volume fraction and disappeared in the ODF after 50% deformation by cold rolling. The typical deformation texture components {112}<111> Copper, {123}<634> S, {110}<112> Brass, and {110}<100> Goss increased with increasing rolling degree. After 50% thickness reduction, the Brass texture component revealed the highest intensity, whereas the S texture component was characterized by the highest volume fraction. At this stage, the texture showed both a well developed β-fiber as well as an α-fiber. Furthermore, the γ-fiber consisting of the {111}<110> E and {111}<112> F texture components did not develop during rolling, whereas the {552}<115> CopperTwin texture component increased in volume fraction after 50% thickness reduction.

### 3.2. Material behavior during annealing

To investigate the material behavior during heat treatment, 50% cold-rolled specimens were isochronally annealed in the temperature range between 500 °C and 900 °C for 1 h. The microstructure and texture evolution during annealing is shown in Figs. 6-9. Up to 600 °C, no recrystallized grains were observed and the deformation-induced microstructural features, such as slip lines, micro shear bands, and deformation twins, were retained (Fig. 6a). At 650 °C, static recrystallization was initiated, and the microstructure consisted of 50% recrystallized grains with a mean grain size of ~2.5 µm (including annealing twins). The grain structure revealed a bimodal character in the partially recrystallized state containing elongated, non-recrystallized grains and heterogeneously distributed recrystallized grains decorating the prior grain boundaries and triple junctions of the deformed matrix grains, as shown in Fig. 6b (see also the subdivided EBSD maps of only non-RX and RX grains in Fig. 14). Furthermore, deformation twins were not only thermally stable during static recovery, but also remained in the microstructure up to high recrystallized volume fractions, as exemplified in Fig. 7. Annealing at 700 °C for 1 h caused almost complete recrystallization (93% recrystallized volume) and slight coarsening of the recrystallized grains to a mean diameter of ~2.8 µm (Fig. 6c). Annealing at higher temperatures facilitated complete transformation of the deformed grains and grain growth up to a mean diameter of ~22.8 µm (Fig. 6d).

---

[‡] Randomly oriented grains are defined as grains with orientation not belonging to any of the ideal rolling texture components. In the calculation, the volume fraction of all ideal texture components is summed up and subtracted from a total volume fraction of 1. The remaining volume fraction is assigned to randomly oriented grains.

At annealing temperatures up to 600 °C, where no recrystallized grains were observed, the character of the texture remained unchanged (Fig. 8). In this temperature range, the volume fraction of most of the main deformation texture components increased, whereas the fraction of randomly oriented grains decreased, indicating slight texture strengthening (Fig. 9). With progressing recrystallization, the intensity and volume fraction of the rolling texture components Copper, S, Brass, Goss, and CopperTwin decreased. In contrast, the volume fraction of the Cube and {110}<110> RotatedGoss texture components increased as well as that of randomly oriented grains and the evolution of the {110}<111> A texture component did not show a clear trend. Additionally, a weak complete α-fiber and an almost complete fiber at (0°-90°, 60°, 45°) were formed. Based on the ODFs and the evolution of the volume fractions of the texture components, the texture after recrystallization can be described as a weak retained rolling texture that was further randomized during grain growth.

3.3. *Material behavior during tensile testing*

The results of tensile tests of the alloy after 50% cold rolling and additional annealing are displayed in Fig. 10. As shown by the engineering stress-strain curves (Fig. 10a) and the true stress-strain curves (Fig. 10b), the cold-rolled material was characterized by the highest yield and ultimate tensile strength but lowest total elongation. With increasing annealing temperature, the strength decreased and the formability increased. Whereas the slope of the stress-strain curves of the different conditions are as expected, the work-hardening rate-true strain curves show a rather surprising trend. The cold-rolled as well as the cold-rolled and at 600 °C annealed specimens work-hardened only slightly and were characterized by a strongly declining work-hardening rate. After annealing at 900 °C the work-hardening rate of the alloy showed a much smoother but also steady decline from ~2500 MPa at a strain of 0.035 to ~900 MPa at a strain of 0.33. By contrast, the specimens annealed at 650 °C and 700 °C revealed a local minimum of the strain-hardening rate at a strain of ~0.03 followed by an increase of the slope. After reaching a local maximum at strain of ~0.038 (650 °C) and ~0.045 (700 °C), the work-hardening rates of both curves dropped again until reaching the UTS. The local maximum and the overall work-hardening rate of the material annealed at 700 °C were at a higher level as compared to the specimens annealed at 650 °C.

The microstructures of the 50% cold-rolled and of the additionally at 650 °C annealed material strained until fracture are shown in Figs. 11 and 12. The 50% cold-rolled specimens revealed a high density of deformation twins that were finer in thickness (~20 nm) as compared to the deformation twins induced during cold rolling (Fig. 11a). Furthermore, secondary twinning within twins formed during cold rolling was also observed (Fig. 11b). Pronounced deformation twinning was also observed within the non-RX grains in the material annealed at 650 °C (Fig. 12a and c). However, plastic deformation of the recrystallized grains was dominantly accommodated by dislocation slip (Fig. 12b and c) and deformation-induced twinning was very scarce within these grains (Fig. 12d).

## 4. Discussion

### 4.1. *Material behavior during cold rolling/microstructure-texture correlation*

Decisive for the development of the microstructure and texture of materials is the stacking fault energy (SFE), which determines the possibility and characteristic of dislocation splitting. At room temperature, the Cantor alloy is known to possess a low SFE that has been estimated to be in the range between 18.3 and 27.3 mJ/m$^2$ [25, 30]. In general, the behavior of a material can be associated to this property and the Cantor alloy is not an exception as will be discussed in this and following sections. To begin with, the relationship between the microstructure and texture will be considered. In the initial stages of deformation (10% - 25%), the grains deformed primarily by dislocation slip as substantiated by numerous slip lines in the microstructure (Fig. 1a and b). However, at 25% thickness reduction deformation twinning was observed, as an additional deformation mechanism (Fig. 1b), which is in well agreement with the observations by Stepanov et al. [21]. At higher degrees of deformation, the volume fraction of the deformation twins increased significantly (Fig. 1d), whereas pronounced alignment of twin-matrix lamellae with the rolling plane and shear banding have not been observed. However, a detailed analysis using TEM revealed that the microstructure consisted mainly of dislocation cell structures and comparatively thick deformation twins (Fig. 2). The evolution of the texture reflects these microstructural changes. The initial texture before room-temperature plastic deformation was characterized by a large volume fraction (~60%) of random orientations. Non-random components occupied respectively less than 10% of the volume (Fig. 5). At low deformation degrees, deformation proceeded mainly by planar slip. This caused an increase of the Cu, S and Brass texture components, a substantial decrease of random orientations, and thus the formation of a Cu-type texture. At higher deformation (50%), as a consequence of the activated deformation mechanisms, a weak rolling texture was formed. This texture can be described as a transition texture between the Cu-type and Brass-type, as it consist of increased fractions of the β-fiber texture components Cu, S, and Brass and of the α-fiber texture components Brass and Goss [41, 42]. However, the weak intensity of the Cu compared to the Brass texture components, where the ODF maximum was located, clearly indicates the tendency to form the Brass-type texture. In addition, the onset of twinning originated a considerable and sudden increase of the volume fractions of CuT orientations. Owing to the relatively low maximum underwent deformation (50%) by the material no substantial rotation of Brass to Goss orientations was observed, as usually found in low-SFE materials at high rolling degrees [43-45]. Furthermore, the agglomeration and collapse of twin-matrix lamellae leading to the formation of shear bands that are created as a necessity to accommodate more plastic deformation occurred infrequently (Fig. 2 c and d). Correspondingly, the γ-fiber, which is a result of shear banding and strong latent hardening due to pronounced planarity of dislocation slip and rotation of twin-matrix lamellae [46-48], did not develop. The comparatively low maximum rolling degree of 50% was also the reason for the medium volume fraction of the Brass texture component and of the weak overall development of the rolling texture. According to studies on 70/30 brass [49, 50], very high rolling degrees (> 95%) are required to achieve the level of saturation of shear bands in the microstructure. This level is necessary to return to more homogeneous deformation processes, which in turn facilitates the formation of a strong Brass texture component as a final stable state [51].

In contrast to other low-SFE materials, such as Brass and austenitic steels, fcc-based HEAs do not show pronounced deformation-induced twinning during rolling at low degrees of cold rolling and tensile testing. Under both loading conditions the contribution of deformation twinning to the plastic strain is less in the HEAs [21, 26] as compared to Brass [52] and austenitic steels [53-56]. For instance, in a 50% cold-rolled high-manganese TWIP steel the Cu and S texture components decreased in volume fraction, whereas the CuT texture component was already saturated and the γ-fiber was formed [41, 57]. This development of the these texture components indicated the formation of a Brass-type texture, which has not yet been observed in the 50% cold-rolled HEA. As a consequence, the transition from Cu-type to Brass-type texture is retarded and shifted towards higher degrees of rolling reduction in the Cantor alloy.

### 4.2. *Material behavior during annealing/microstructure-texture correlation*

During heat treatment, the microstructure and the texture underwent substantial changes. The macrotexture showed at initial times of annealing a sharpening of the rolling texture components (Figs. 8 and 9) and an increase of the texture index (Fig. 13). This intensification of the texture is associated with the recovery of the material, which leads to less scattering of X-rays during XRD measurements and thus, to slight texture sharpening. After prolonged times and upon recrystallization, a weakening of the texture, i.e. a decrease of the texture index, accompanied with a characteristic decrease of the microhardness was observed (Fig. 13). However, the main rolling texture components were retained. In general, this behavior has also been observed in fcc materials with similar SFE , for instance, TWIP steels that have been subjected to low and medium rolling degrees [58]. Similar features have also been observed in copper alloys [59].

To understand the texture evolution during recrystallization, the active nucleation mechanisms must be revealed. In a previous study on low-SFE TWIP steel subjected to cold rolling and annealing, we identified nucleation of recrystallized grains at grain boundaries and triple junctions due to strain-induced grain boundary migration to be the preferred mechanism [38]. In this case, the orientations of nuclei were defined by the orientation of subgrains/dislocation cells in the deformed matrix. The same retainment of the matrix orientation has been observed in the investigated HEA, as evidenced by the nucleus in Fig. 12c. In order to reveal further the correlation between microstructure and texture evolution in the present HEA, the EBSD mappings of the partially recrystallized states after annealing at 650 °C and 700 °C (Fig. 6b and c) were analyzed in detail, as shown in Fig. 14. Based on the in-grain orientation deviation (using the GROD-AO value), the grains within the EBSD mappings have been decomposed into recrystallized and non-recrystallized (deformed) grains. As evinced by the pattern of the recrystallized grains after annealing at 650 °C, new grains were distributed heterogeneously due to their preferred formation at grain boundaries and triple junctions. It is well known that nuclei stemming from these crystal defects tend to have orientations close to the parent grains [60, 61]. For this reason, the deformation texture is conserved after annealing, as can be seen from the comparison of the microtexture ODFs of the recrystallized and non-recrystallized grains after annealing at 650 °C. With progressing recrystallization after annealing at 700 °C the texture was further randomized. This development agrees well with the macrotexture evolution (Figs. 8 and 9). A slight texture randomization may be expected from these mechanisms because the orientations scatter slightly from that of the parent grain owing to their rotation and fragmentation during deformation. However, the main mechanism that modifies the texture is the appearance of annealing twins during recrystallization [62]. In contrast to previous works [62-64], where higher order twin chains have been observed, predominantly first order twins with $\sum3$-orientation relationship were detected in the investigated HEA. Since the described nucleation at grain boundaries and triple junctions facilitated the formation of new grains with orientations belonging to the rolling texture components located on the α-fiber (Fig. 14, state 650 °C), the annealing twins must have a $\sum3$-orientation relationship with these texture components. In dependence on Refs. [38, 65], the orientations that develop due to twinning of the ideal α-fiber texture components Goss, Goss/Brass and Brass have been computed and are displayed in Fig. 15. Obviously, first order twinning within Goss-, Goss/Brass-, and Brass-oriented nuclei results in the formation of a complete α-fiber due to the formation of the P, A, and rotated Goss texture components. In addition, a fiber parallel to the α-fiber located at (0°-90°, 30°, 45°) is expected to develop. Indeed, formation of a complete α-fiber containing a strengthened rotated Goss texture component and the evolution of a weak, incomplete fiber parallel to the α-fiber was observed (Figs. 8 and 9). Furthermore, annealing twinning was not only initiated in nuclei with α-fiber orientations, but in all orientations present in the spectrum of nuclei, which led to a wide scatter of new orientations and further texture randomization [66, 67]. This is in agreement with the increasing volume fraction of randomly oriented grains with increasing annealing temperature (Fig. 9).

To analyze further the influence of grain-boundary nucleation and annealing twinning on the texture evolution in the investigated HEA, CA simulations of the recrystallization process were performed. In Fig. 16, the results of these simulations are shown. For the CA simulations, twin orientations stemming from the 50% deformed matrix grains and orientations stemming from grain-boundary nucleation of the parent grains were first separately (a), (b) and then jointly (c) considered. In Fig. 16a, the texture is dominated by the twin orientations of the texture components formed after deformation. The rotation corresponds to a $\sum$3-orientation relationship (60°<111>). The resulting texture becomes randomized (Fig. 15) as evinced by the maximum intensity in Fig. 16b owing to the fact that 60°<111> rotations of many orientations do not belong to the main texture components depicted in Fig. 15. By contrast, the texture shown in Fig. 16b is characterized by the retention of slightly weaker deformation texture components. As aforementioned, nuclei originating at grain boundaries tend to have orientation close to the parent grains. Once both mechanisms are active in the simulations (Fig. 16c), a good agreement with the experimentally measured orientation spectrum (Fig. 16d) of the nuclei is found only if most orientations stem from the twinned volume. For the orientation distribution function shown in Fig. 16c, the volume fraction occupied by twin-orientations (Fig. 16b) corresponds to 66%. Because some orientations are found in both datasets (grain boundary nuclei and twin orientations) a comparison with experiment is difficult as discriminating both mechanisms in experiment is extremely laborious. Nevertheless, the simulations clearly point out that the main mechanism for the randomization of the texture is the formation of twin orientations during annealing, whereas grain boundary nucleation facilitates a retention of the deformation texture components. It is noted that some of the orientation found after recrystallization can appear only because of twinning. This is the case of the α-fiber that forms by the twin rotation of the Goss, Goss/Brass and Brass orientations as shown in Fig. 15.

### 4.3. *Deformation behavior during tensile testing*

Annealing of the 50% cold-rolled samples had a significant effect on the deformation behavior during tensile testing. To begin with, during annealing at 600 °C static recovery processes caused a slight decrease in strength and an increase in the total elongation (Fig. 10a). The formation of micro shear bands during the previous cold rolling (Fig. 1c) indicates a local saturation of the dislocation density leading to heterogeneous, localized deformation. Hence, the 50% cold-rolled and the at 600 °C annealed specimens were hardly capable to store further dislocations, resulting in strongly decreasing work-hardening rates (Fig. 10b) regardless of active deformation twinning (Fig. 11). With the onset of primary recrystallization and with increasing annealing temperature and recrystallized volume fraction, the yield and ultimate tensile strength decreased further whereas the total elongation increased.

Surprisingly, the partially recrystallized states (650 °C/1 h and 700 °C/1 h) showed two inflection points in the slopes of their respective work-hardening rate-true strain curves (Fig. 10b). A detailed illustration of this region is given in the work-hardening rate-true stress curve (Kocks-Mecking plot) and the work-hardening rate-true strain curve of the 700 °C/1 h material in Fig. 17. According to Gutierrez and Raabe, such curves (Fig. 17a and b) can be divided into five stages, A to E [54]. In the case of the investigated HEA, stage A is comparable to the work-hardening stage III found in high-SFE metals, where the dynamic recovery processes facilitate a strongly, continuously decreasing work-hardening rate [68]. The transition from stage A to stage C, i.e. stage B, is very narrow. This regime is usually associated with the formation of dislocation substructures such as dislocation cells and highly dense dislocation walls [54, 69]. These substructures act as precursors for the nucleation of deformation twins at grain boundaries [70]. In stage C, an additional deformation mechanism must be activated that strongly reduces the mean free path of dislocations and facilitates an increase of the work-hardening rate. However, stage C of the investigated HEA was also narrow (strain range from 0.03 to 0.04 in Fig. 17b) and much less pronounced compared to materials with pronounced deformation twinning [54, 71]. At a higher strain, the reduced ability to store further dislocations in the microstructure led to another inflection point in the work-hardening rate-true stress/strain curve and then to a continuously decreasing slope (stages D and E). As evidenced by TEM investigation, deformation twinning was mainly initiated in the non-recrystallized grains during tension, while deformation twins were hardly created in the recrystallized grains (Figs. 11 and 12). This deformation behavior caused the peculiar slope of the work-hardening rate of the partially recrystallized material. On the one hand, the initially high dislocation density within the non-recrystallized grains enabled locally to reach the high resolved shear stress for activation of mechanically induced twinning and served as precursor for nucleation of deformation twins. Hence, stage B was narrow because twin nuclei preexisted in the deformed grains. On the other hand, plastic deformation of the partially recrystallized material was mainly accommodated by dislocation slip in the recrystallized grains, whereas the hard non-recrystallized grains underwent minimal deformation, only in order to co-rotate and guarantee form continuity. Therefore, the limited contribution of deformation twinning in the non-recrystallized grains accounted for the narrow stage C. The main contribution to the high work-hardenability of the partially recrystallized and fully recrystallized material stemmed undoubtedly from the activation of multiple slip systems, the strong planarity of dislocation slip, and the related low ability for dislocation cross-slip, as usually observed in fcc-based HEAs [26, 72, 73]. This behavior is in agreement with other alloys that deform primarily by planar dislocation slip, e.g. [48], and explains the highest work-hardenability of the fully recrystallized material.

## 5. Conclusions

The equiatomic CoCrFeMnNi high-entropy alloy has been subjected to cold rolling and subsequent annealing at varying annealing temperature. The influence of the physical mechanisms activated during deformation and recrystallization on the microstructure evolution have been identified and correlated with the texture evolution. A specific emphasis was put on the effect of deformation and annealing twinning. The following conclusions can be drawn:

- During cold rolling a transition from the Cu-type to the Brass-type texture occurs at intermediate rolling degrees. This behavior is associated with the formation of deformation twins and the related strong latent hardening. However, the texture transition is shifted towards higher degrees of cold rolling due to the limited twinning activity, as compared to other low-SFE alloys.

- The microstructure and texture evolution during recrystallization is dominated by grain-boundary nucleation and annealing twinning. Nuclei formed at grain boundaries retained the texture components of the rolling texture, whereas annealing twins caused both the formation of prominent twin orientations as well as an overall texture randomization.

- During deformation by tensile testing deformation twinning was mainly initiated within non-recrystallized grains. Due to the limited contribution of non-recrystallized grains to the overall plasticity of partially recrystallized material, the stage C work-hardening regime was very narrow in the investigated HEA as compared to other low-SFE alloys.

- The high work-hardenability was promoted by the strong planarity of dislocation slip and the reduced tendency for dynamic recovery in the recrystallized grains rather than by deformation twinning.

## Acknowledgements

The authors acknowledge gratefully the financial support of the Deutsche Forschungsgemeinschaft (DFG) within the Collaborative Research Center (SFB) 761 "Steel - ab initio; quantum mechanics guided design of new Fe based materials". The authors are thankful for the help of Arndt Ziemons, Marie Luise Köhler, Mehran Afshar, Christian Schnatterer, Johannes Lohmar, and Stefan Senge with carrying out the experiments.

**Tables**

Table 1: Chemical composition of the investigated steel.

| Element | Co | Cr | Fe | Mn | Ni | O (ppm) |
|---------|------|------|------|------|------|---------|
| (wt.%) | 21.8 | 17.9 | 20.3 | 18.8 | 21.0 | 135 |

Table 2: Definition of texture components illustrated in Fig. 3.

| Component | Symbol | Miller indices | Euler angles ($\varphi_1$, $\Phi$, $\varphi_2$) | Fiber |
|-----------|--------|----------------|------------------------------------------------|-------|
| Brass (B) |  | {110}<112> | (55, 90, 45) | α, β |
| Goss (G) |  | {110}<100> | (90, 90, 45) | α, τ |
| Goss/Brass (G/B) |  | {110}<115> | (74, 90, 45) | α |
| Rotated Goss (RtG) |  | {110}<110> | (0, 90, 45) | α |
| A |  | {110}<111> | (35, 90, 45) | α |
| P |  | {011}<211> | (30, 90, 45) | α |
| Cube (C) |  | {001}<100> | (45, 0, 45) | / |
| E |  | {111}<110> | (0/ 60, 55, 45) | γ |
| F |  | {111}<112> | (30/ 90, 55, 45) | γ |
| Copper (Cu) |  | {112}<111> | (90, 35, 45) | β, τ |
| CopperTwin (CuT) |  | {552}<115> | (90, 74, 45) | τ |
| S |  | {123}<634> | (59, 37, 63) | β |
| α-fiber | <110> parallel to ND | | | |
| β-fiber | <110> tilted 60° from ND towards RD | | | |
| γ-fiber | <111> parallel ND | | | |
| τ-fiber | <110> parallel TD | | | |

**Figures**

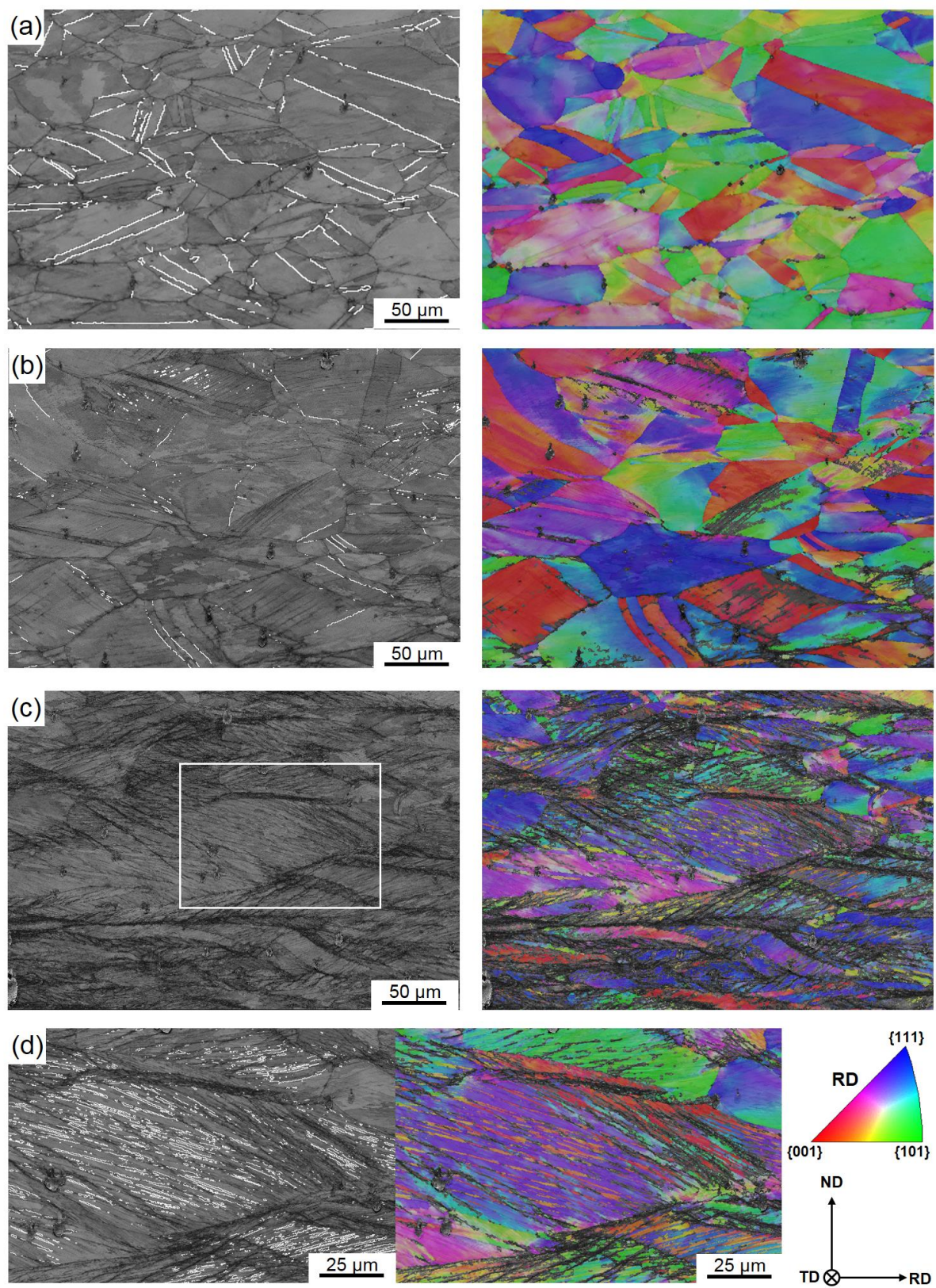

Figure 1: EBSD-band contrast (BC) maps (left) and BC maps overlaid by inverse pole figure (IPF) maps (right) of the material after (a) 10% cold rolling, (b) 25% cold rolling, and (c) 50% cold rolling. Substantial formation of deformation twins was observed after 50% cold rolling as shown in the inset of (c) in (d). White lines denote ∑3 (60°<111>) grain boundaries.

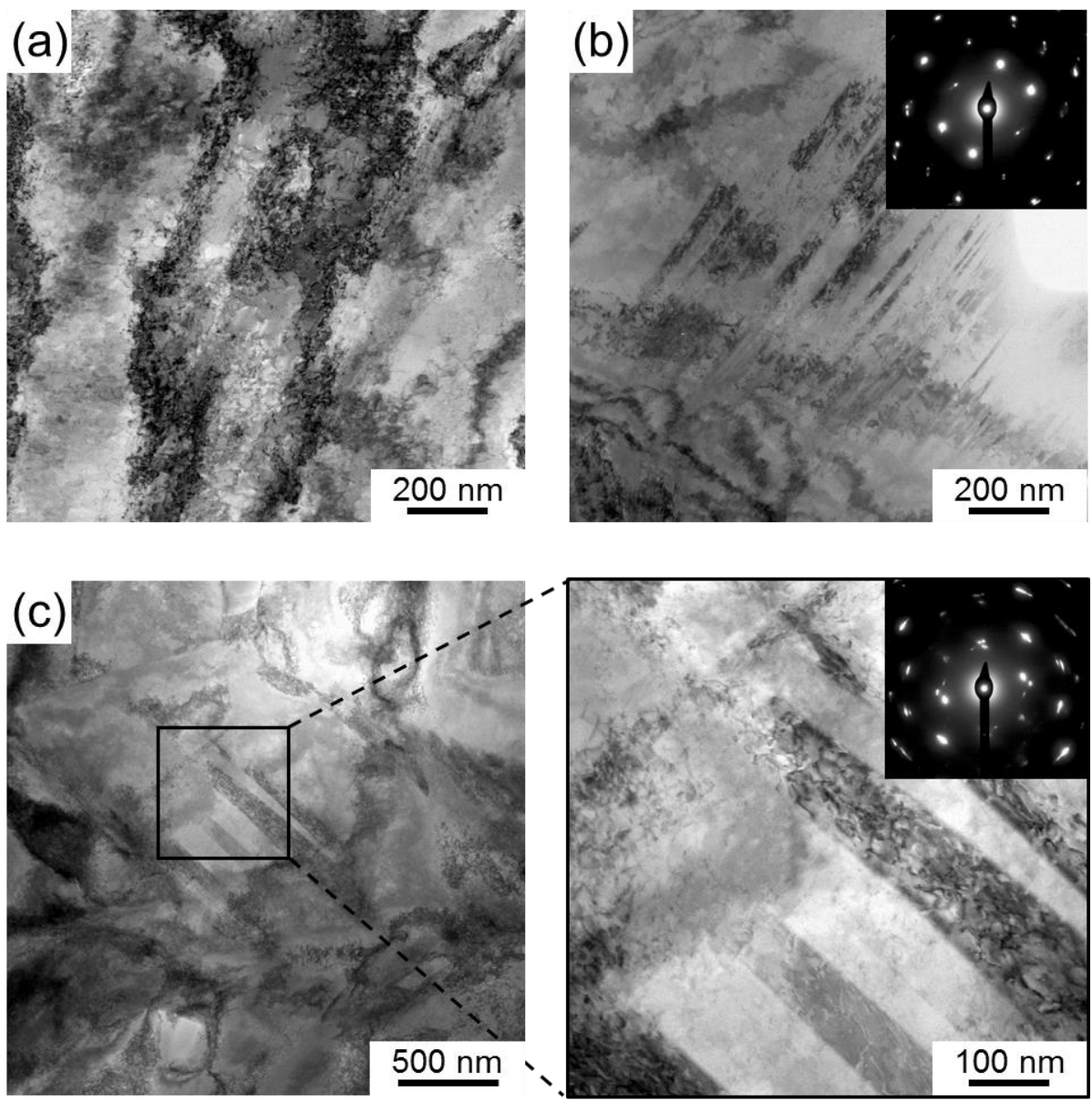

Figure 2: TEM bright field (BF) images of the material after 50% cold rolling. (a) The deformed microstructure is dominated by the occurrence of dislocation cells. (b) & (c) In addition, thick deformation twins are observed. These twins were found to contain a high density of dislocations (inset in (c)). The selected area electron diffraction (SAED) pattern reveal the twin orientation relationship.

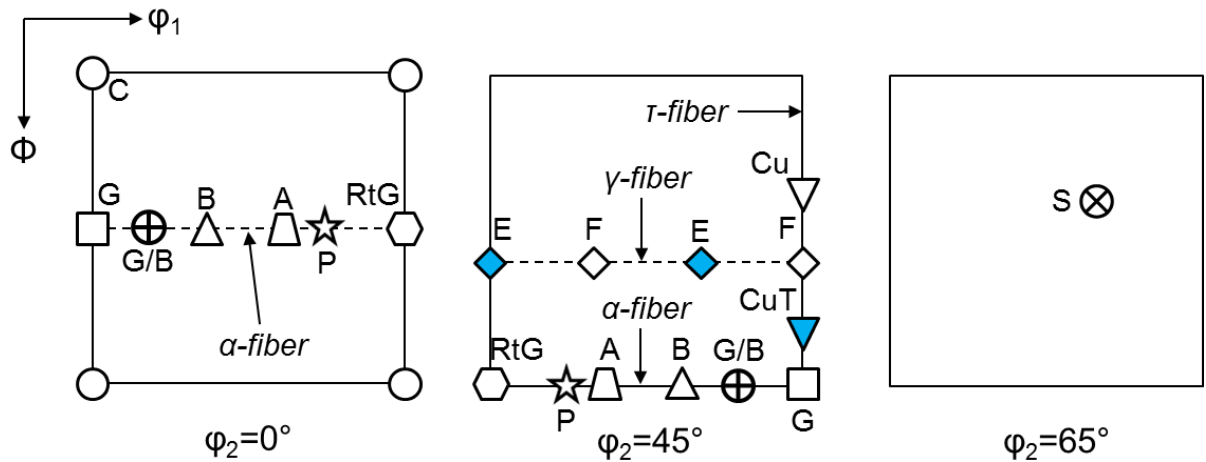

Figure 3: Schematic illustration of the ideal texture components and fibers formed in the alloy investigated, ODF sections at $\phi_2 = 0°$, 45°, and 65°. The texture components are defined in Table 2.

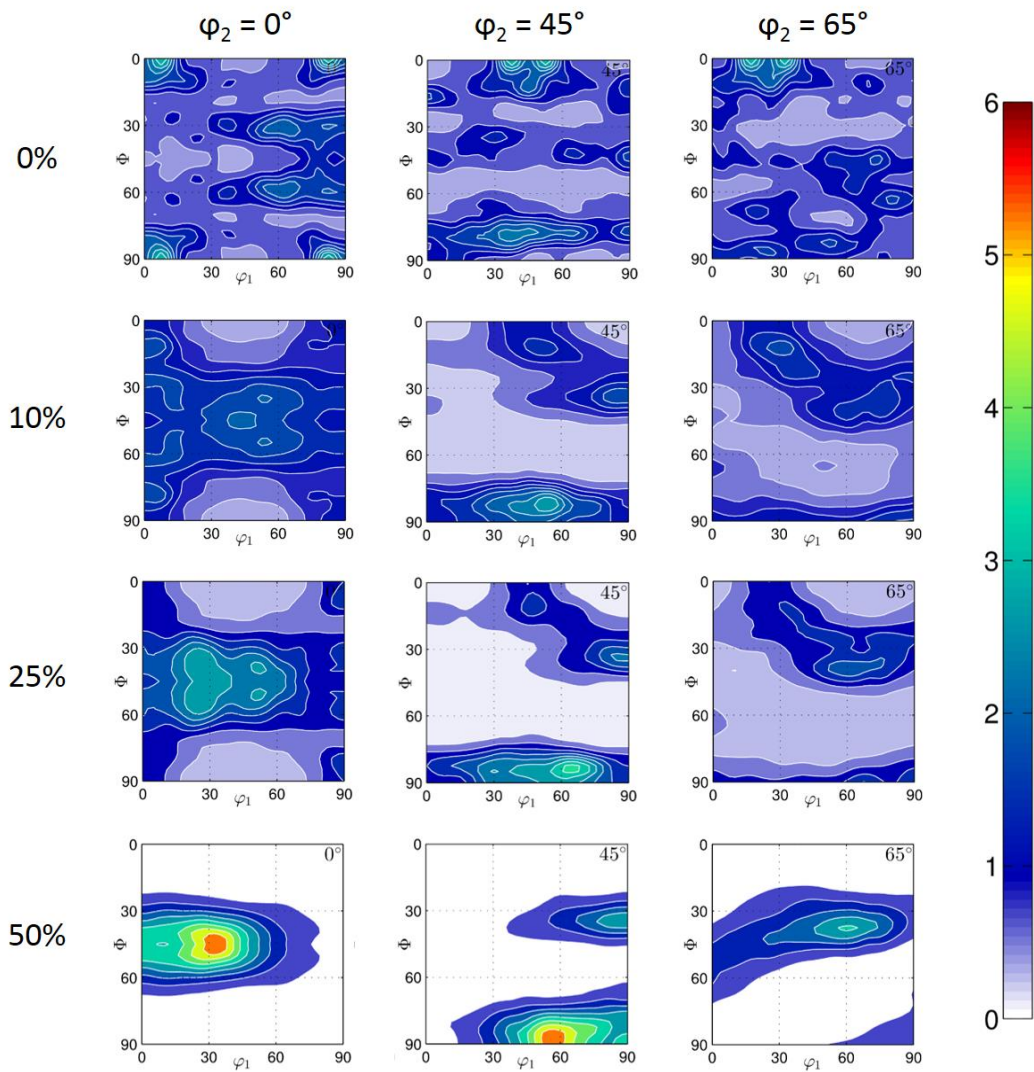

Figure 4: Texture evolution of the investigated steel during cold rolling, ODF sections at $\phi_2 = 0°$, 45°, and 65°.

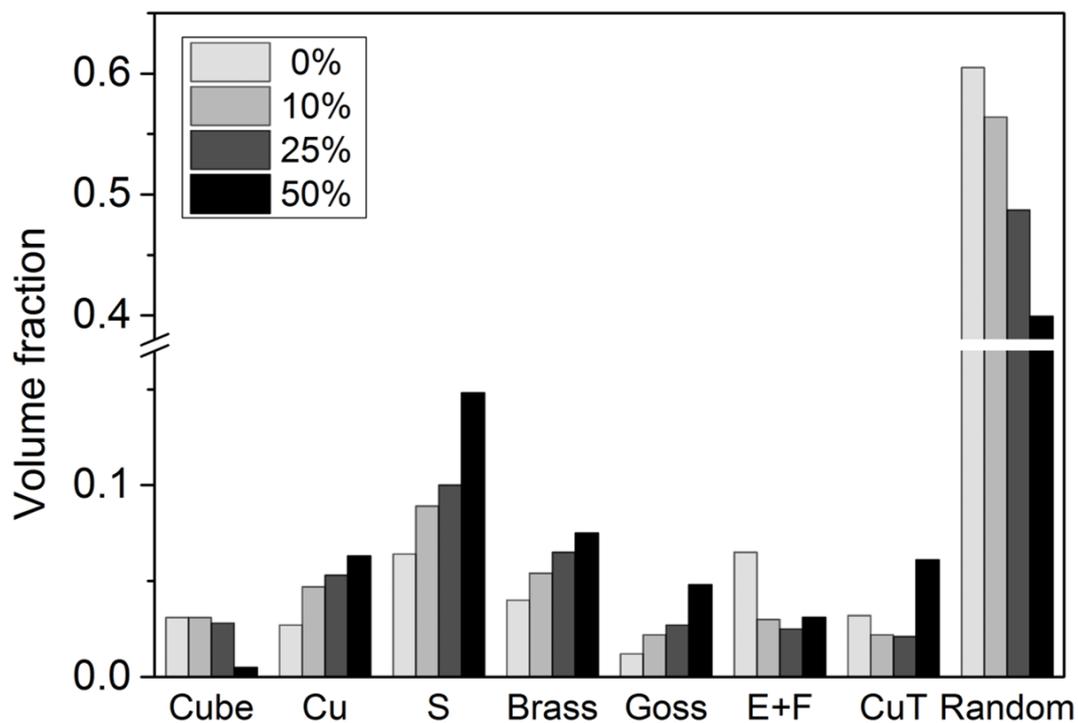

Figure 5: Volume fractions of the main texture components developed during cold rolling.

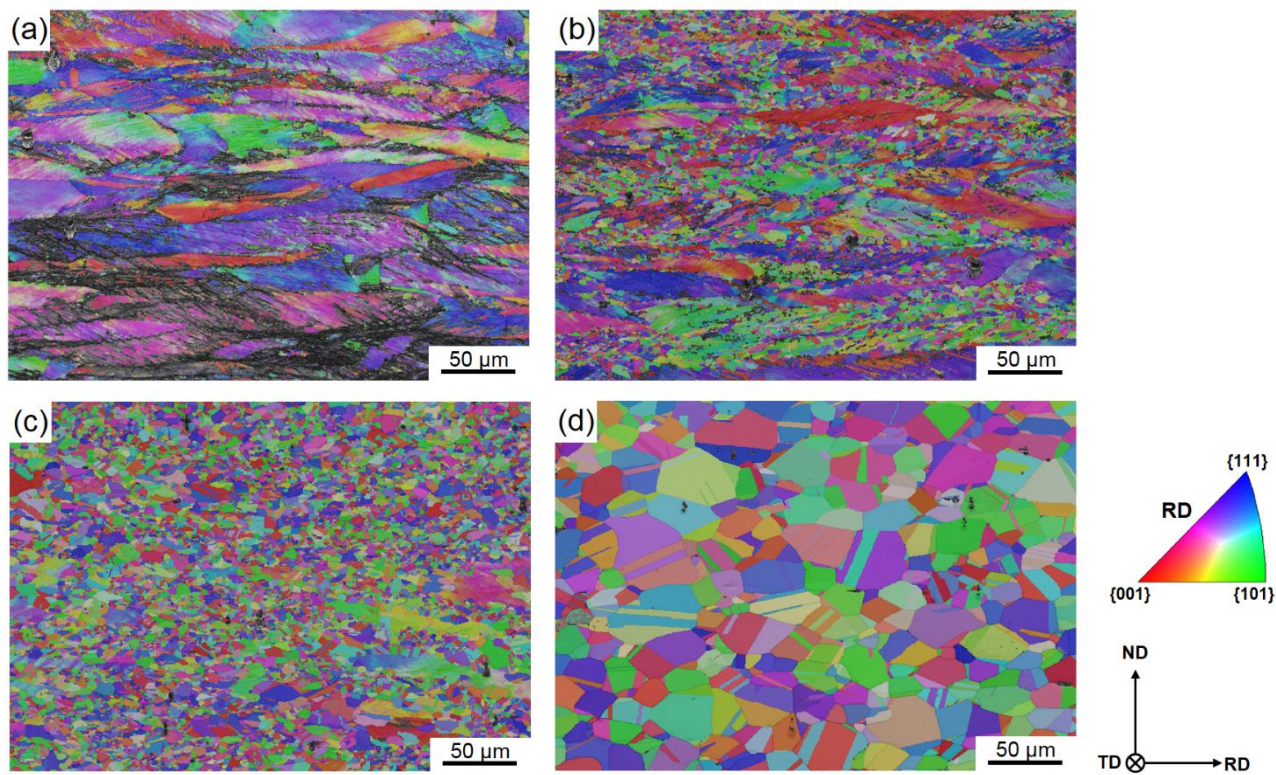

Figure 6: EBSD-BC maps overlaid by IPF maps of the material after 50% cold rolling and annealing for 1 h at (a) 600 °C, (b) 650 °C, (c) 700 °C, and (d) 900 °C.

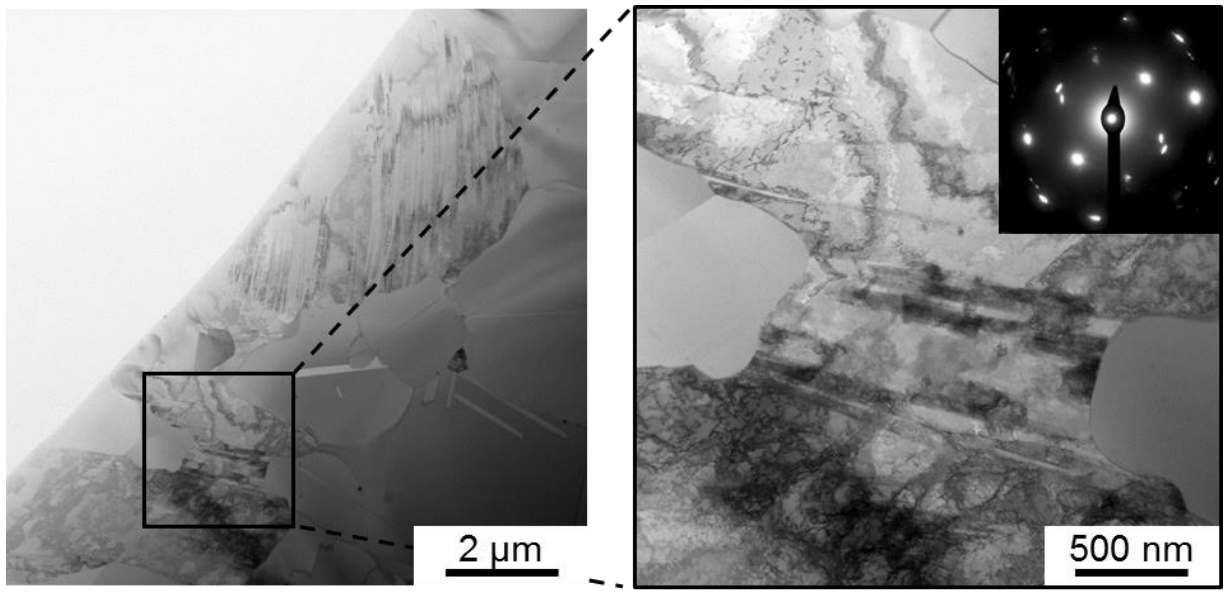

Figure 7: TEM BF image of the material after 50% cold rolling and annealing at 650 °C for 1 h. The inset highlights a non-recrystallized matrix grain containing deformation twins surrounded by recrystallization nuclei. The SAED pattern was taken from the deformed matrix and confirms the twin orientation relationship.

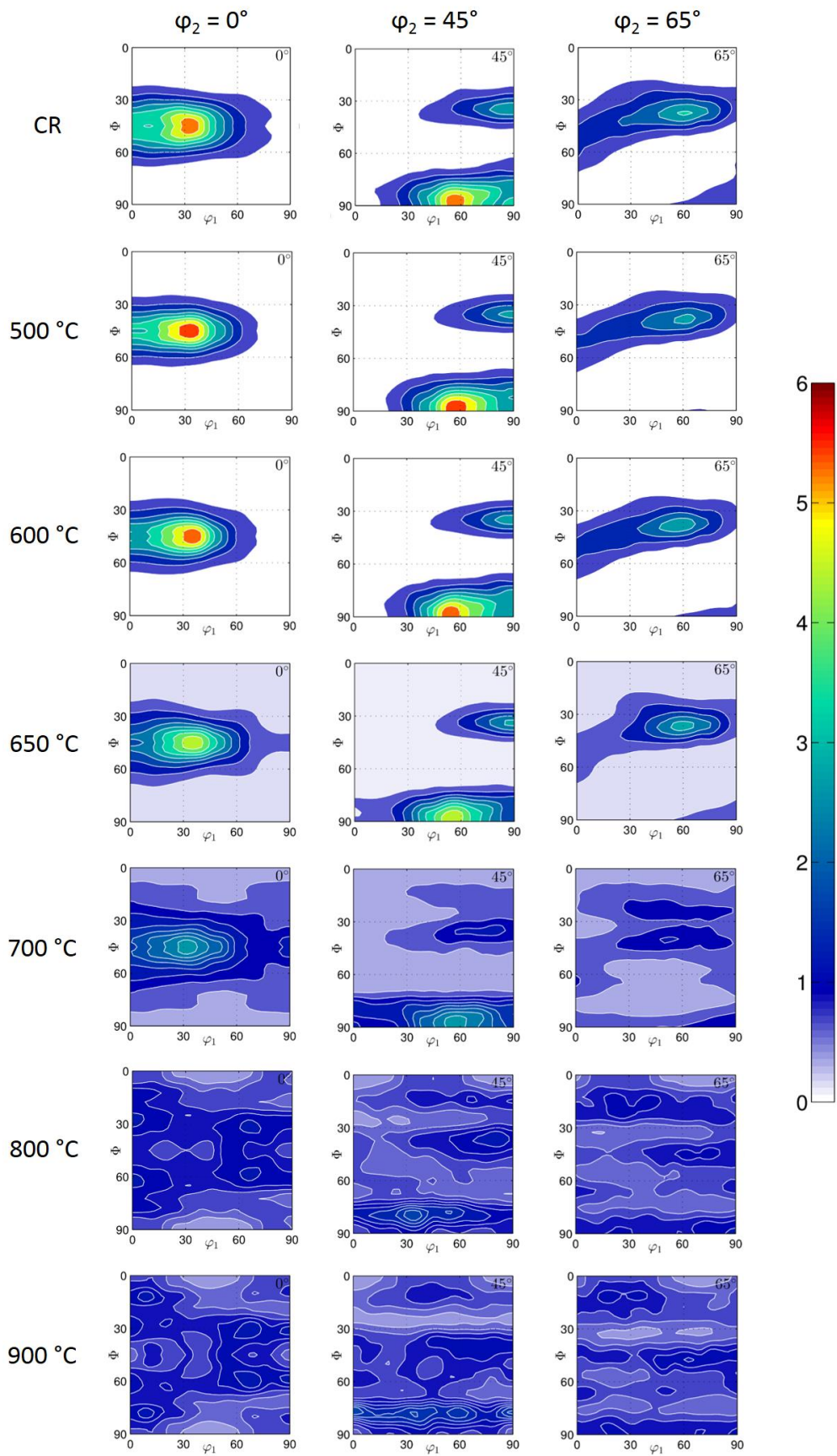

Figure 8: Texture evolution of the investigated steel during annealing, ODF sections at $\phi_2 = 0°$, 45°, and 65°.

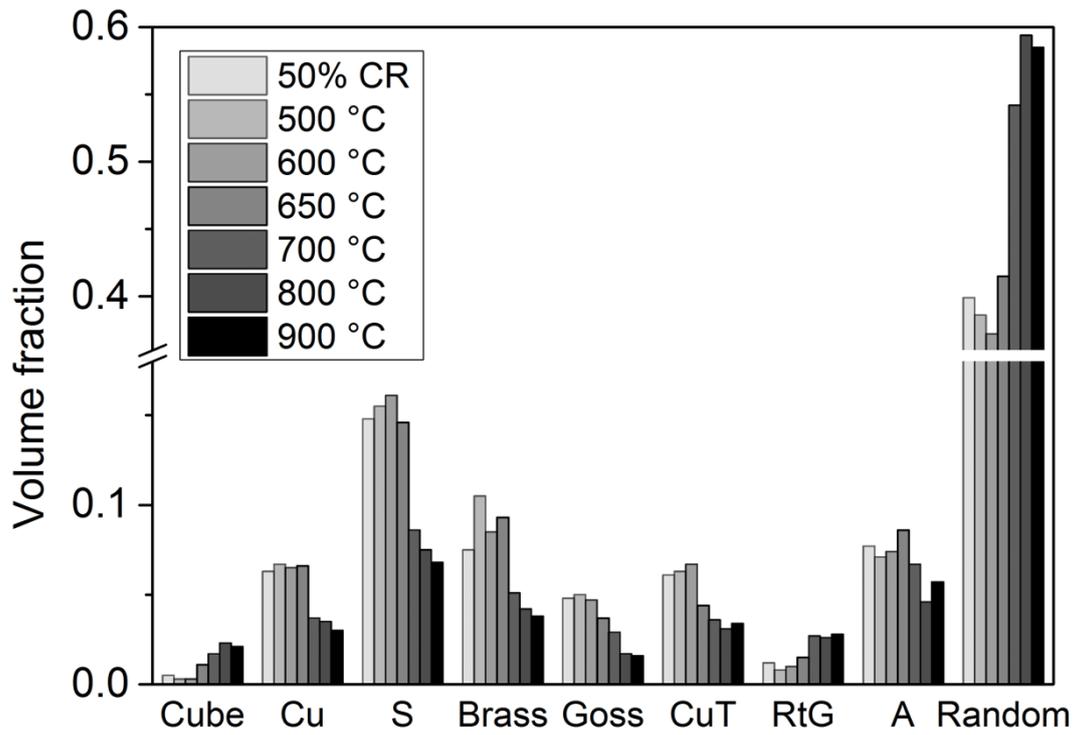

Figure 9: Volume fractions of the main texture components developed during annealing of the 50% cold-rolled material.

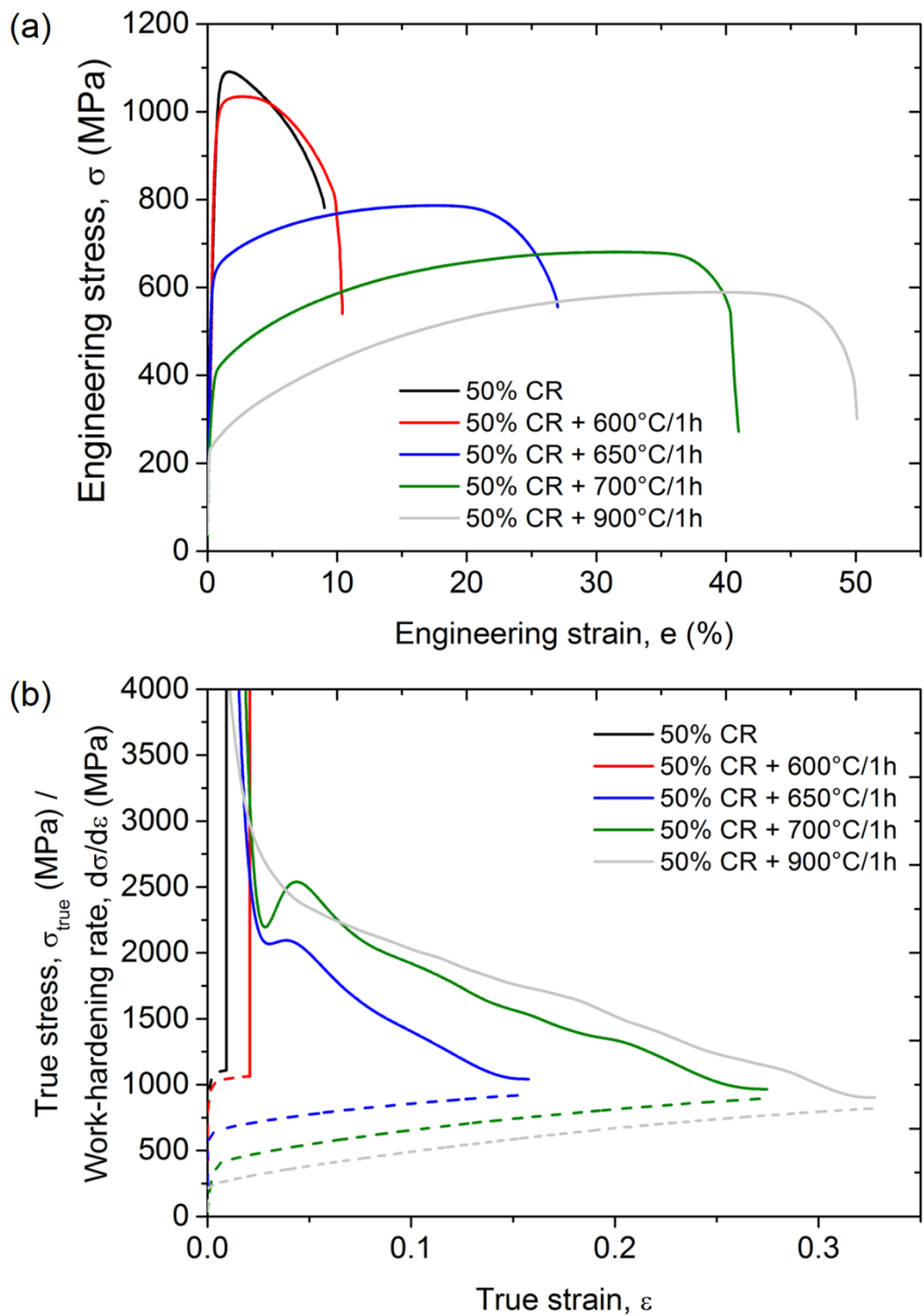

Figure 10: (a) Engineering stress–strain curves and (b) true stress–true strain curves (dotted lines) and work-hardening rate–true strain curves (complete lines) of the investigated alloy.

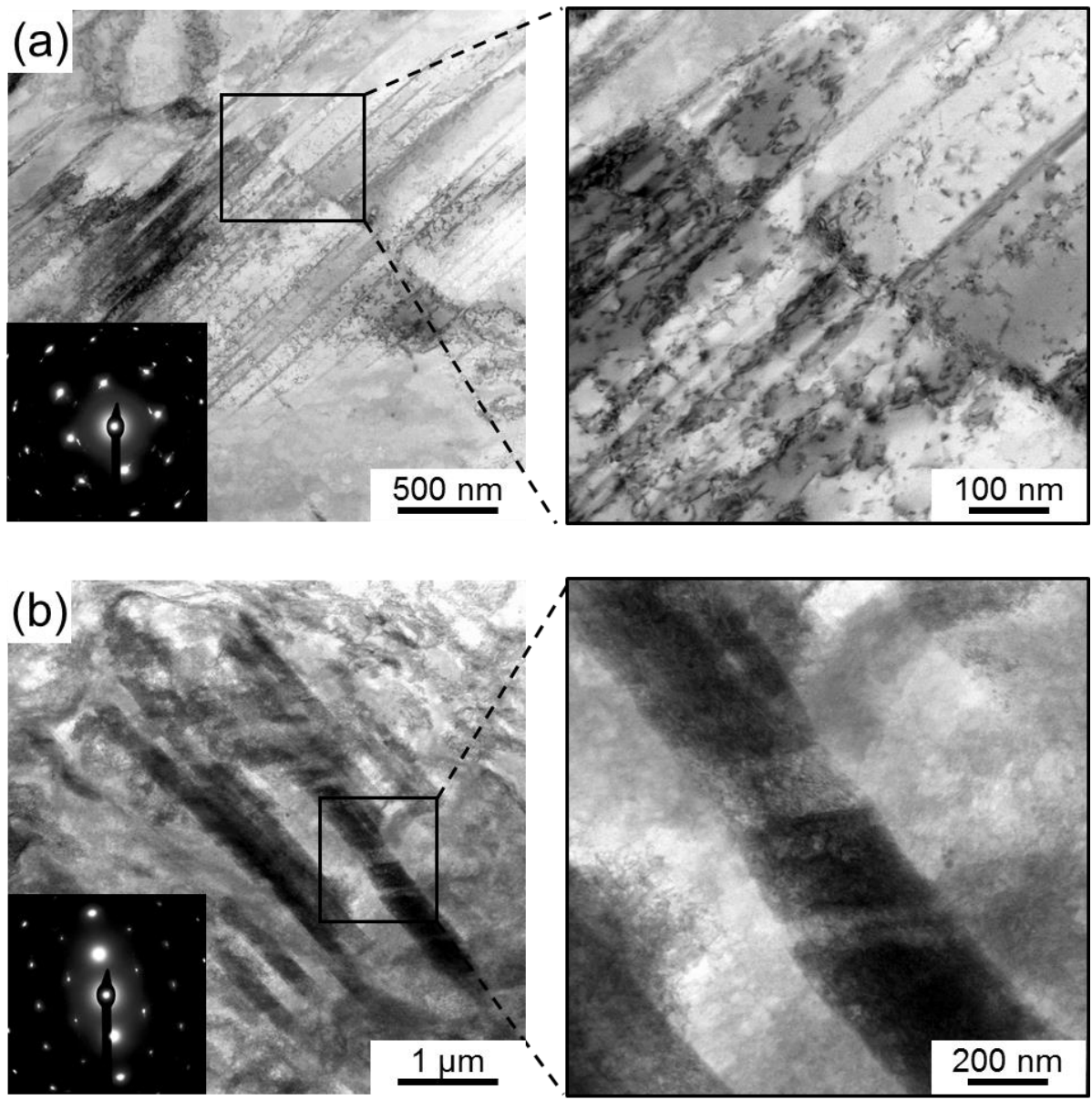

Figure 11: TEM BF images of the material after 50% cold rolling and tensile testing until fracture. (a) A high density of fine deformation twins is formed during tensile testing. (b) Thicker deformation twins induced by cold rolling undergo secondary twinning during further deformation by tensile testing, as highlighted by the inset of (b).

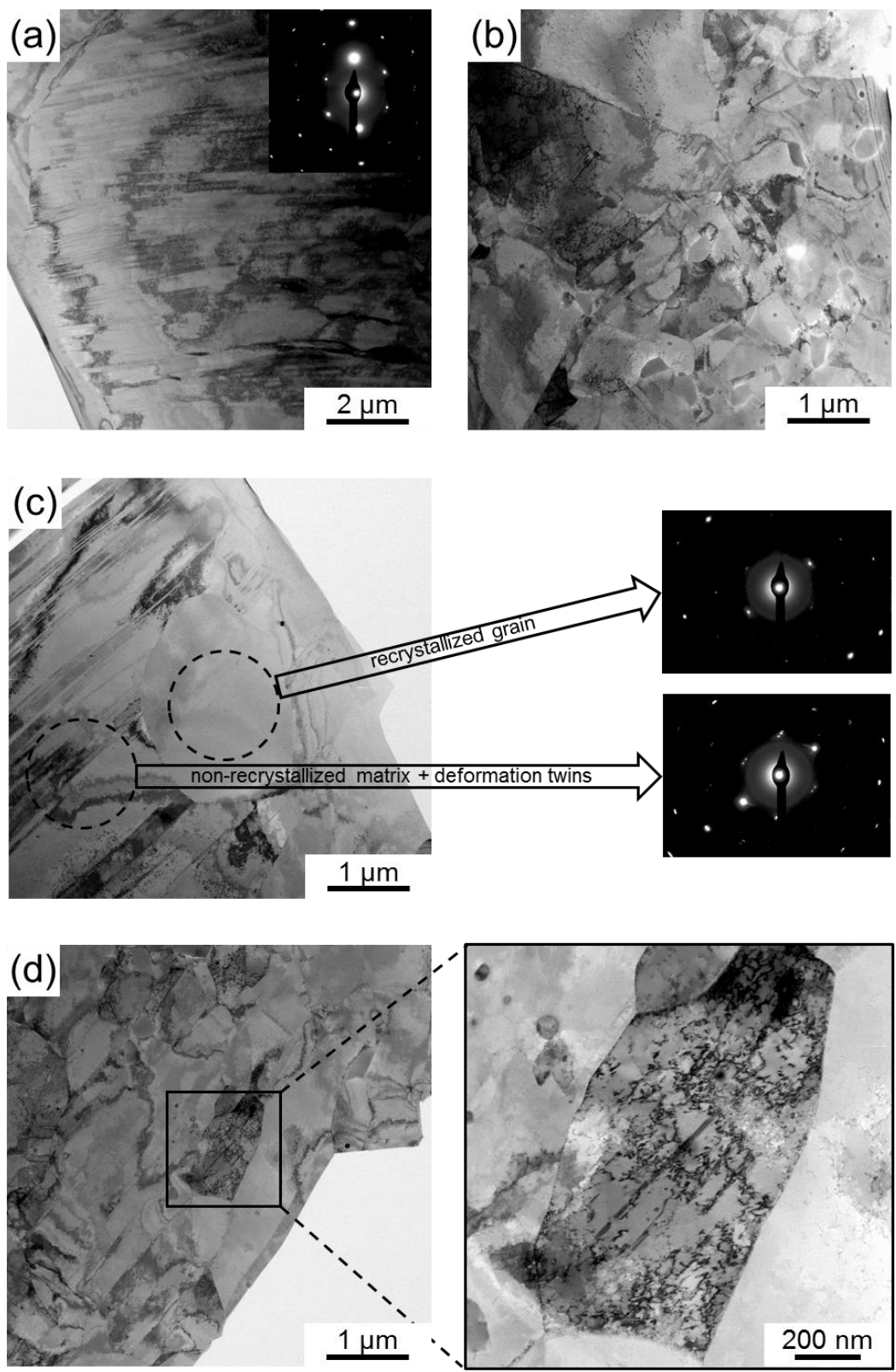

Figure 12: TEM BF images of the material after 50% cold rolling, annealing at 650 °C for 1 h and tensile testing until fracture. (a) A high density of fine deformation twins formed in the non-recrystallized matrix during tension. (b) & (c) In contrast, the recrystallized grains did not undergo deformation twinning. The SAED pattern in (c) reveal that the recrystallized grain has the same orientation as the non-recrystallized matrix. Deformation twinning in recrystallized grains was only observed sparsely, as shown in the inset of (d).

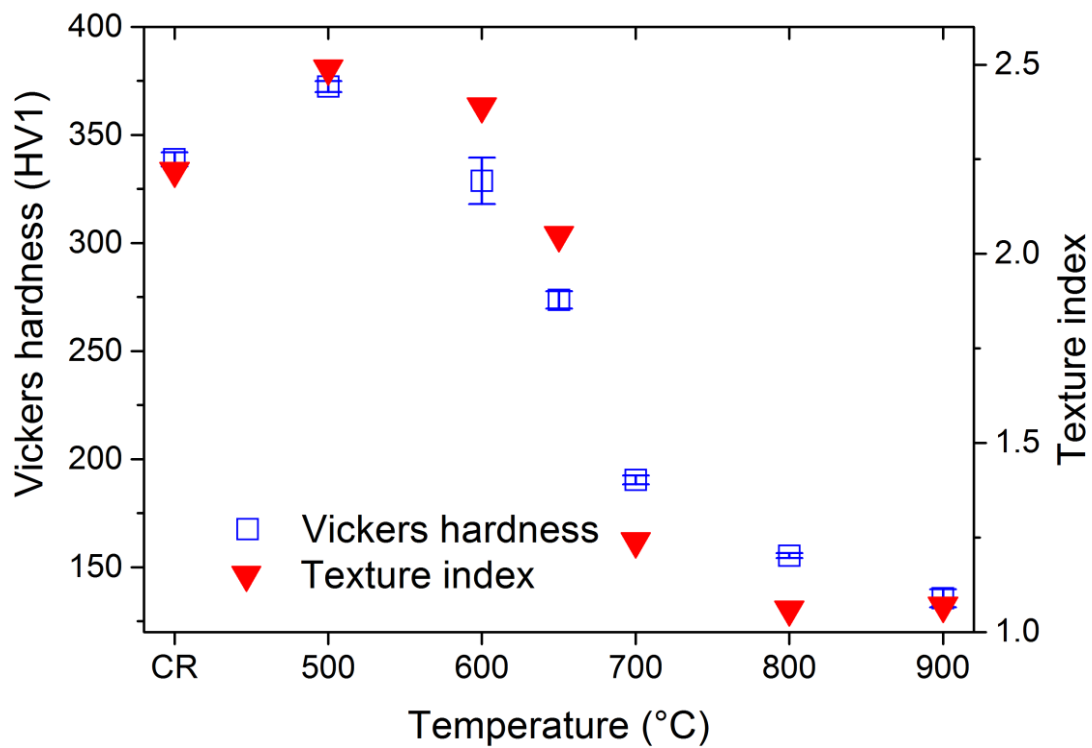

Figure 13: Microhardness and texture-index evolution after 50% cold rolling and annealing for 1 h at varying temperature.

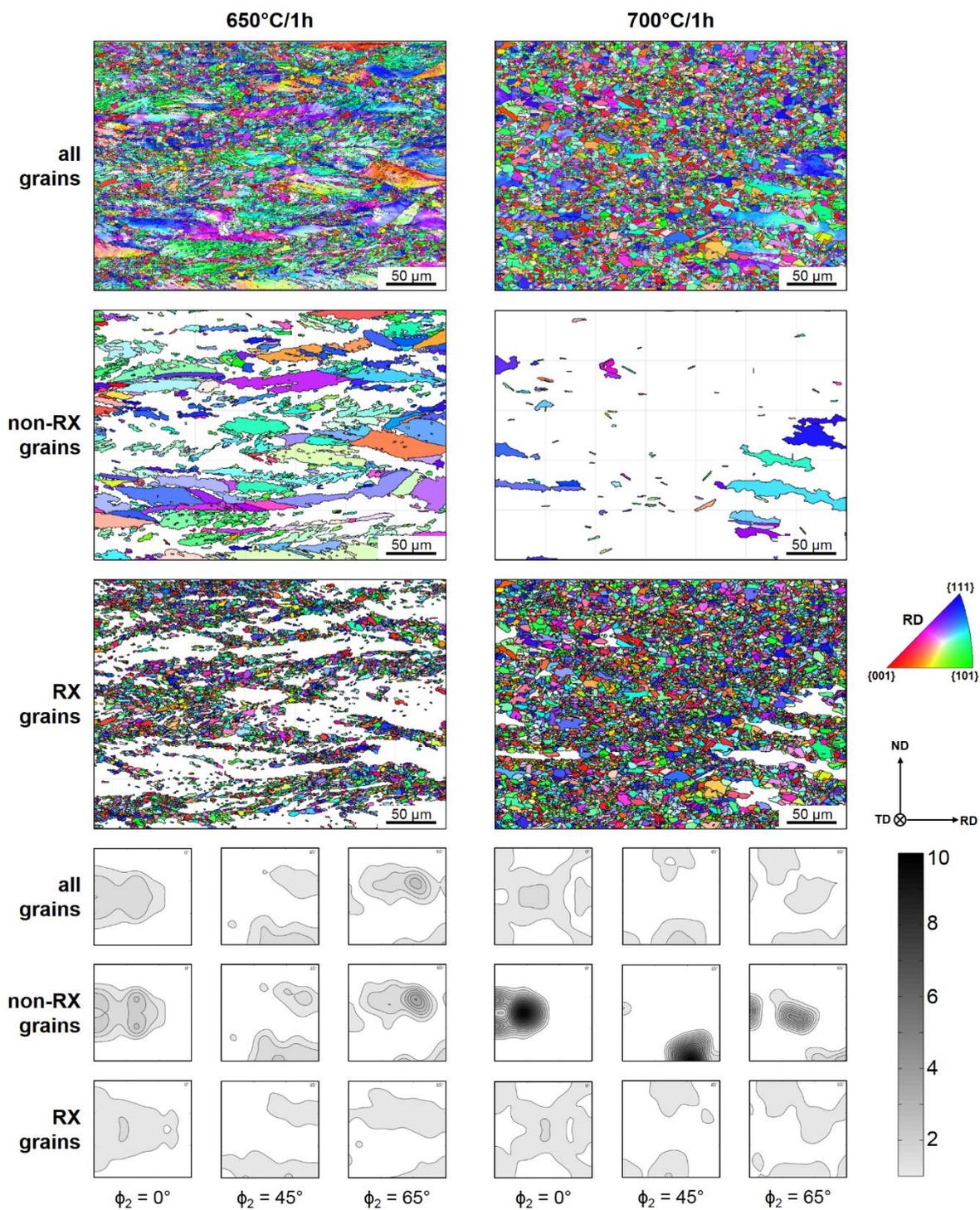

Figure 14: EBSD maps with IPF color-coding of the material after 50% cold rolling and annealing at 650 °C and 700 °C, respectively. The EBSD maps containing all grains were subdivided into maps of non-RX grains and RX grains based on the GROD-AO value of each grain. The corresponding microtexture ODFs are presented for constant $\varphi_2$ angles.

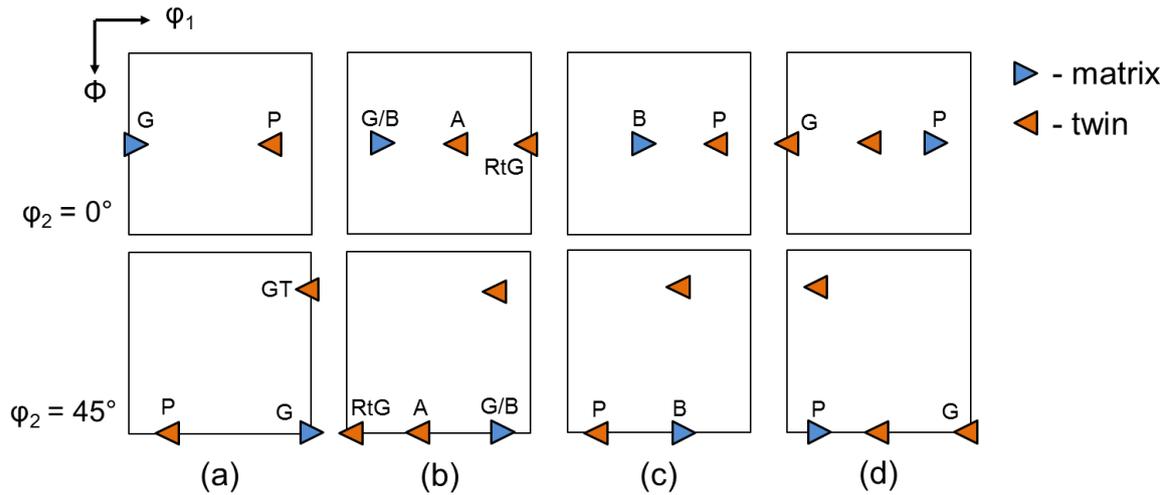

Figure 15: Orientations of first order twins (60°<111> orientation relation) with respect to the matrix texture components (a) Goss, (b) Goss/Brass, (c) Brass, and (d) P.

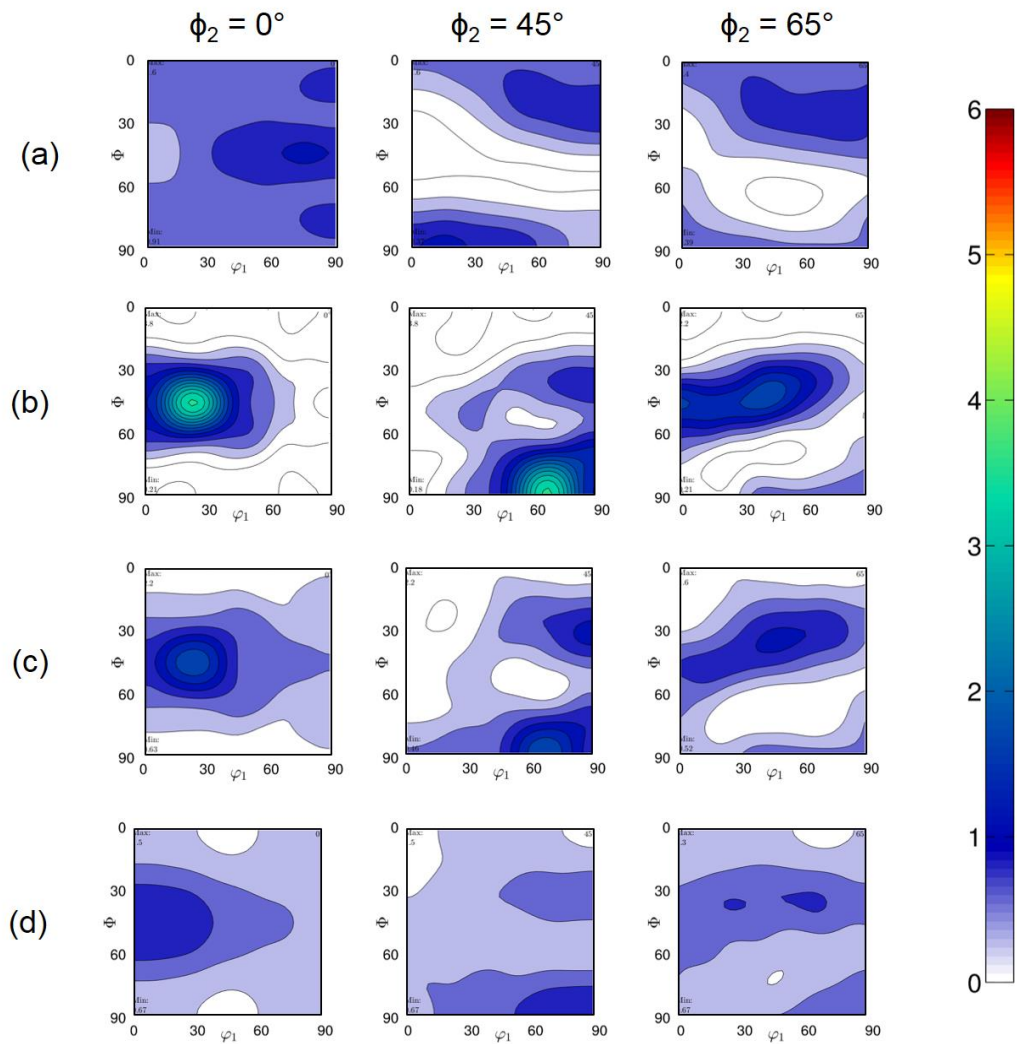

Figure 16: Simulated textures by the cellular automaton method considering (a) only orientations of primary twins formed by 60°<111> rotation of the 50% deformed grains (condition 50% in Fig. 4), (b) only orientations of grains nucleating at grain boundaries; nuclei

are considered to form with orientations close to the 50% deformed grains within a scatter of 5°, (c) consideration grain boundary nucleation and annealing twinning, (d) considering only the orientations extracted as recrystallized grains from experiments (condition 650°C in Fig. 14), ODF sections at $\phi_2 = 0°$, 45°, and 65°.

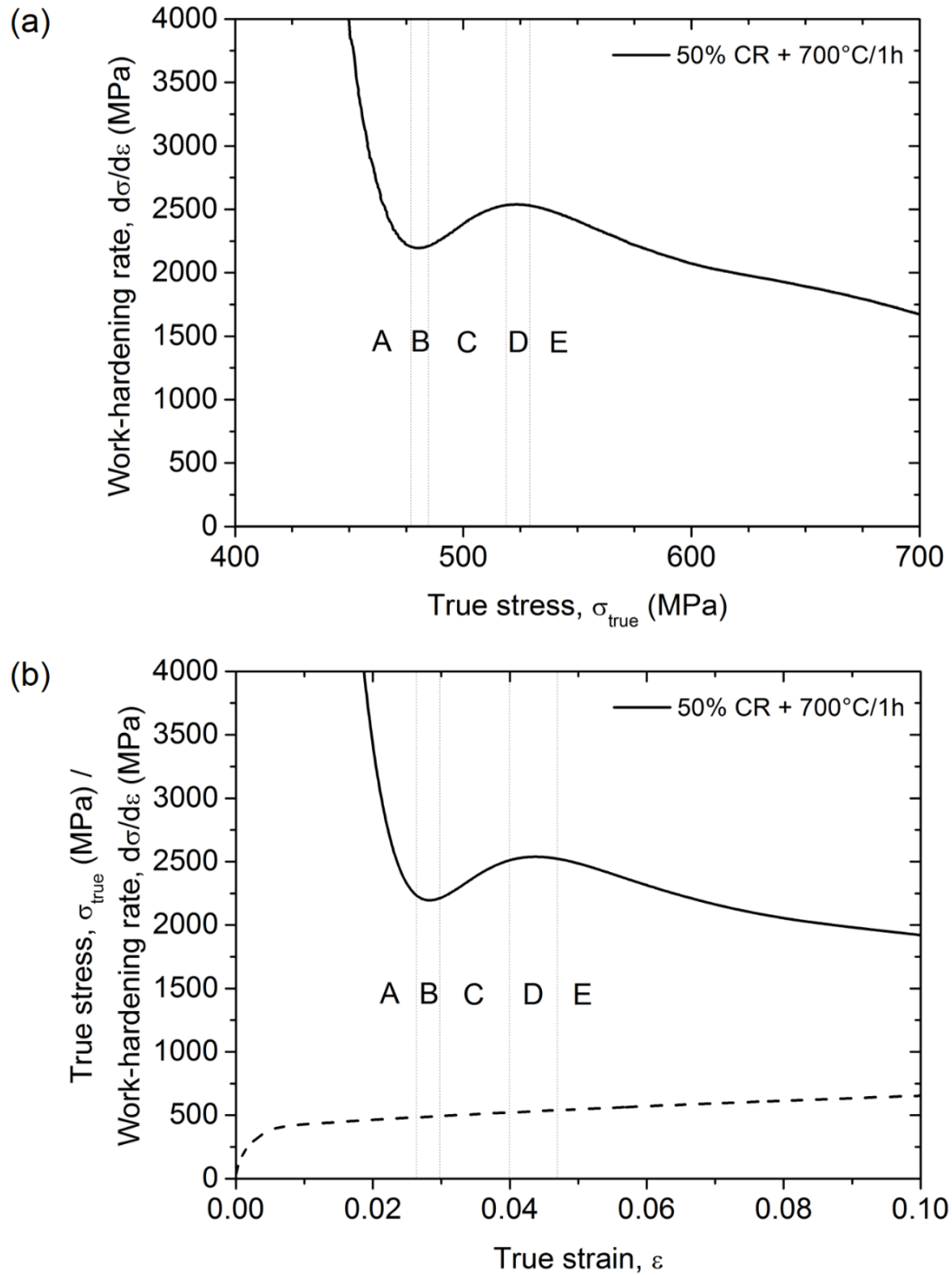

Figure 17: (a) Work-hardening rate-true stress curve and (b) trues stress-strain (dotted line) and work-hardening rate-true strain curves (complete line) of the 50% cold-rolled and at 700 °C annealed material. A to E indicate different work-hardening regimes. The curves in (b) are shown up to a true plastic strain of 0.1. The complete curves are illustrated in Fig. 10.